%% file: main.tex
\documentclass[aps,prb,twocolumn,showpacs,a4paper]{revtex4}
\usepackage{amsmath,amssymb,amsfonts}
\usepackage{graphicx}

\newcommand{\ham}{\mathcal{H}}
\newcommand{\saf}[1]{\textsf{#1}}
\newcommand{\nocc}[2]{n_{\saf{#1},{#2}}}

\begin{document}

\title
{
  Thermodynamic anomalies in a lattice model of water:
  Solvation properties
}
\author{M. Pretti and C. Buzano}
\affiliation
{
  Istituto Nazionale per la Fisica della Materia (INFM)
  and Dipartimento di Fisica, \\ Politecnico di Torino,
  Corso Duca degli Abruzzi 24, I-10129 Torino, Italy
}
\date{\today}
\begin{abstract}
We investigate a lattice-fluid model of water, defined on a
3-dimensional body-centered cubic lattice. Model molecules possess
a tetrahedral symmetry, with four equivalent bonding arms. The
model is similar to the one proposed by Roberts and Debenedetti
[J. Chem. Phys. {\bf 105}, 658 (1996)], simplified by removing
distinction between ``donors'' and ``acceptors''. We focus on
solvation properties, mainly as far as an ideally inert
(hydrophobic) solute is concerned. As in our previous analysis,
devoted to neat water [J. Chem. Phys. {\bf 121}, 11856 (2004)], we
make use of a generalized first order approximation on a
tetrahedral cluster. We show that the model exhibits quite a
coherent picture of water thermodynamics, reproducing
qualitatively several anomalous properties observed both in pure
water and in solutions of hydrophobic solutes. As far as
supercooled liquid water is concerned, the model is consistent
with the second critical point scenario.
\end{abstract}


\maketitle

\section{Introduction}

From the experimental point of view, water is known to exhibit
several thermodynamic
anomalies~\cite{EisenbergKauzmann1969,Franks1982,Stanley2003}.
Contrary to most fluids, at ordinary pressure the solid phase
(ice) is less dense than the liquid phase. The latter displays a
temperature of maximum density at constant pressure, while both
isothermal compressibility and isobaric heat capacity display a
minimum as a function of temperature. Moreover, heat capacity is
on average much larger than usual. Anomalous properties of neat
water have been studied for long, but much interest has been
devoted also to unusual properties of water as a solvent, in
particular for nonpolar (hydrophobic) chemical
species~\cite{FrankEvans1945,BenNaim1980,DillScience1990,SouthallDillHaymet2002}.
Insertion of a nonpolar solute molecule in water is characterized
by positive solvation Gibbs free energy (it is thermodynamically
unfavored), negative solvation enthalpy (it is energetically {\em
favored}), negative solvation entropy (it has an ordering effect),
and large positive solvation heat
capacity~\cite{CrovettoFernandez-PriniJapas1982,BenNaim1987}. More
precisely, for prototype hydrophobic species (that is, for
instance, noble gases), solvation entropies and enthalpies, which
are negative at room temperature, increase upon increasing
temperature, and eventually become positive. These properties
define the so-called hydrophobic effect, whose importance in
biological processes, such as protein folding, has been emphasized
in the latest years~\cite{Dill1990}.

From the theoretical point of  view, one can relate the anomalous
properties of neat water to the formation of a large amount of
hydrogen bonds, because of peculiar features of water
molecules~\cite{Stanley1998,Poole1994}. The same physics is
believed to underly the hydrophobic
effect~\cite{FrankEvans1945,BenNaim1980,Stillinger1980}, but a
comprehensive theory which explains all of these phenomena has not
been developed yet. A possible way of investigation consists of
computer
simulations~\cite{MahoneyJorgensen2000,Stanley2002,Paschek2004},
based on very detailed (but still phenomenological) interaction
potentials. In this way, quite a high level of accuracy in
describing water thermodynamics has been achieved. Nevertheless,
simulations are generally limited by the large computational
effort required, while microscopic physical mechanisms are
sometimes hidden by a large number of model details. A
complementary approach involves investigations of simplified
models~\cite{SouthallDillHaymet2002,AshbaughTruskettDebenedetti2002,WidomBhimalapuramKoga2003,BruscoliniCasetti2001pre}.
Although quantitative accuracy is sometimes poor, this approach
usually allows more detailed analysis, in a wide range of
thermodynamic conditions, with relatively low computational cost.
One of these attempts is based on the application of
scaled-particle theory~\cite{ReissFrischLebowitz1965} to
hydrophobic hydration~\cite{Lee1991}. A recent descendant of
scaled-particle theory is the information theory approach by Pratt
and coworkers~\cite{Hummer_et_al1996}, based on previous knowledge
of water properties, such as the pair correlation function, which
can be obtained by experiments or by simulations. According to the
cited studies, the hydrophobic effect would result mostly from
small size of water molecules, and not from water structuring by
the solute, as in the classical view~\cite{FrankEvans1945}. Such
effect, though existing, would be scarcely relevant for a
description of the hydrophobic effect. The simplified molecular
thermodynamic theory of
Ref.~\onlinecite{AshbaughTruskettDebenedetti2002} is basically in
agreement with this conclusion. Different theories stress that the
large positive heat capacity variation, observed upon insertion of
apolar solutes into water, can only arise from cooperativity, that
is, from induced ordering of water molecules, so that a theory of
the hydrophobic effect should be based on a description of this
phenomenon. This position is supported by the results of the
2-dimensional ``Mercedes Benz'' model, first introduced by
Ben-Naim in 1971~\cite{BenNaim1971}, and extensively investigated
by Dill and coworkers in the latest
years~\cite{SouthallDillHaymet2002}. Contrary to the previously
mentioned approaches, the Mercedes Benz model, though involving
high simplifications, is based on well defined microscopic
interactions, that is, on an energy function, without previous
knowledge of water properties. One important reason to do so is
the need of modelling water in a computationally convenient way,
in order to investigations on complex systems such as
biomolecules, for which water plays a key role. An even more
simplified
approach~\cite{BesselingLyklema1997,SharmaKumar1998,WidomBhimalapuramKoga2003},
which nevertheless can in principle satisfy this criterion, relies
on the long standing tradition of lattice fluid models. As far as
neat water is concerned, several different models, both in
2~\cite{BellLavisI1970,BellLavisII1970,Lavis1973,LavisChristou1979,HuckabyHanna1987,BuzanoDestefanisPelizzolaPretti2004,BalladaresBarbosa2004,DeoliveiraBarbosa2005}
and 3
dimensions~\cite{Bell1972,BellSalt1976,LavisChristou1977,MeijerKikuchiVanRoyen1982,SastrySciortinoStanley1993jcp,BesselingLyklema1994,RobertsDebenedetti1996,PrettiBuzano2004}
have been investigated, some of which are variations of the early
model proposed by Bell in 1972~\cite{Bell1972}. One of them is the
3-dimensional model by Roberts and
Debenedetti~\cite{RobertsDebenedetti1996,RobertsPanagiotopoulosDebenedetti1996,RobertsKarayiannakisDebenedetti1998},
defined on the body-centered cubic lattice. In this model, water
molecules possess four bonding arms (2 donors and 2 acceptors)
arranged in a tetrahedral symmetry. Hydrogen (H) bond formation
requires that 2 nearest neighbor molecules point respectively a
donor and an acceptor towards each other. A number of nonbonding
configurations is allowed, to account for H bond directionality.
Bond weakening occurs (the bond energy is increased of some
fraction) whenever a third molecule is placed near a formed bond.
The latter feature basically mimics the fact that too closely
packed water molecules disfavor H bonding. Let us notice that,
while bonding properties are equivalent to those of the Bell
model~\cite{Bell1972}, the weakening criterion is different. The
Roberts-Debenedetti model is quite appealing in that it has been
shown to predict not only some of real water thermodynamic
anomalies, such as the temperature of maximum density, but also a
liquid-liquid phase separation in the supercooled region, and a
second critical point. Nevertheless, the distinction between
donors and acceptors is likely to be not so crucial to the physics
of water. Therefore, in a previous paper~\cite{PrettiBuzano2004},
we have investigated a simplified version of the model (without
donor/acceptor distinction), showing that the same basic
properties could be reproduced. Here we extend the simplified
model to deal with aqueous solutions, working out solvation
thermodynamics for an inert (apolar) solute. Our purpose is to
verify whether the model is able to reproduce also the main
features of hydrophobicity. This analysis might be interesting
also in view of investigations on mixtures of water with more
complex chemical species, such as polymers.
We shall carry out the analysis by means of a generalized
first-order approximation on a tetrahedral cluster, which has been
verified to be quite accurate for the neat water
model~\cite{PrettiBuzano2004}.

\section{The model}

\input{reticolo.tex}
Let us introduce the model. Molecules are placed on the sites of a
body-centered cubic lattice, whose structure is sketched in
Fig.~\ref{fig:reticolo}. A site may be empty or occupied by a
water molecule ($\saf{w}$) or by a solute molecule ($\saf{s}$). An
attractive potential energy $-\epsilon_{\saf{x} \saf{y}}<0$ is
assigned to any pair of nearest neighbor (NN) sites occupied by
molecules of species $\saf{x},\saf{y}$, where $\saf{x}$ and
$\saf{y}$ can take the values $\saf{w},\saf{s}$. This is the
ordinary Van der Waals contribution. Water molecules possess four
equivalent arms that can form H bonds, arranged in a tetrahedral
symmetry, so that they can point towards 4~out~of~8 NNs of a given
site. There is no distinction between donors and acceptors, so
that a H~bond is formed whenever two NN molecules have a bonding
arm pointing to each other, yielding an energy~$-\eta<0$. It is
easily seen that water molecules have only 2 different
configurations in which they can form H~bonds (see
Fig.~\ref{fig:molecole}). We assume that $w$~more configurations
are allowed, in which water molecules cannot form bonds ($w$~is
related to the bond-breaking entropy). Moreover, we assign an
energy increase~$\eta c_\saf{x}/6$, with $c_\saf{x}\in[0,1]$, for
each of the $6$ sites closest to a formed bond (i.e., 3 out of 6
second neighbors of each bonded molecule), occupied by an
$\saf{x}$ molecule. A bond surrounded by $6$ molecules of
species~$\saf{x}$ contributes an energy~$-\eta (1-c_\saf{x})$. As
far as water molecule are concerned, the weakening parameter
mainly accounts for the fact that H bonds are most favorably
formed when water molecules are located at a certain distance,
larger than the optimal Van der Waals distance. Therefore, if too
many molecules are present, the average distance among them is
decreased, and hydrogen bonds are (on average) weakened. Moreover,
the presence of an external molecule may perturb the electronic
density, resulting in a lowered H~bond strength as well. A
weakening parameter for the solute ($c_\saf{s}$) takes into
account possible perturbation effects for a generic chemical
species, even if in the following we shall mainly consider an
ideally inert solute with $c_\saf{s} = 0$.
\input{molecole.tex}

\input{cactustetraedro.tex}
The hamiltonian of the system can be written as a sum over
irregular tetrahedra, whose vertices lie on 4 different
face-centered cubic sublattices, shown in Fig.~\ref{fig:reticolo}.
One of such tetrahedra is shown in
Fig.~\ref{fig:cactustetraedro}(a). We have
\begin{equation}
  \ham =
  \frac{1}{6} \sum_{\langle \alpha,\beta,\gamma,\delta \rangle}
  \ham_{i_\alpha i_\beta i_\gamma i_\delta}
  ,
  \label{eq:ham}
\end{equation}
where $\ham_{ijkl}$ is a contribution which will be referred to as
tetrahedron hamiltonian, and the subscripts
$i_\alpha,i_\beta,i_\gamma,i_\delta$ label site configurations for
the 4 vertices $\alpha,\beta,\gamma,\delta$, respectively.
Possible site configurations are: ``empty'' ($i=0$), ``bonding
water'' (site occupied by a water molecule in one of the $2$
orientations which can form bonds: $i=1,2$; see
Fig.~\ref{fig:molecole}), ``nonbonding water'' (site occupied by a
water molecule in one of the $w$ orientations which cannot form
bonds: $i=3$), ``solute'' (site occupied by a solute molecule:
$i=4$). Assuming that $(i,j)$, $(j,k)$, $(k,l)$, and $(l,i)$ refer
to NN pair configurations, the tetrahedron hamiltonian reads
\begin{equation}
  \ham_{ijkl} = H_{ijkl} + H_{jkli} + H_{klij} + H_{lijk}
  ,
  \label{eq:tetraham}
\end{equation}
where
\begin{equation}
  H_{ijkl} =
  - \epsilon_\saf{xy} \nocc{x}{i} \nocc{y}{j}
  - \eta h_{ij} \left( 1 - c_\saf{x} \frac{\nocc{x}{k} + \nocc{x}{l}}{2} \right)
  ,
  \label{eq:tetraham2}
\end{equation}
$\nocc{x}{i}$ is an occupation variable for the $\saf{x}$ species,
defined as in Tab.~\ref{tab:configurazioni}, while $h_{ij}$ are
bond variables, defined as $h_{ij}=1$ if the pair configuration
$(i,j)$ forms a H~bond, and $h_{ij}=0$ otherwise. Here and in the
following, we understand that repeated $\saf{x},\saf{y}$ indices
are to be summed over their possible values $\saf{w},\saf{s}$.
\begin{table}
  \caption{
    Site configuration labels~($i$), with corresponding multiplicities~($w_i$)
    and occupation numbers for water ($\nocc{w}{i}=1$) and solute ($\nocc{s}{i}=1$).
  }
  \begin{ruledtabular}
  \begin{tabular}{l|ccccc}
    $i$           & $0$ & $1$ & $2$ & $3$ & $4$ \cr
    \hline
    $w_i$         & $1$ & $1$ & $1$ & $w$ & $1$ \cr
    $\nocc{w}{i}$ & $0$ & $1$ & $1$ & $1$ & $0$ \cr
    $\nocc{s}{i}$ & $0$ & $0$ & $0$ & $0$ & $1$
  \end{tabular}
  \end{ruledtabular}
  \label{tab:configurazioni}
\end{table}
Let us also assume that $i,j,k,l$ (in this order) denote
configurations of sites placed on, say, $A,B,C,D$ sublattices
respectively. If $A,B,C,D$ are defined as in
Fig.~\ref{fig:reticolo}, we can define $h_{ij}=1$ if $i=1$ and
$j=2$, and $h_{ij}=0$ otherwise. With the above assumptions, the
tetrahedron hamiltonian is independent of the orientation, that is
of the arrangement of $A,B,C,D$ on its vertices. Let us notice
that both Van der Waals ($-\epsilon_\saf{xy} \nocc{x}{i}
\nocc{y}{j}$) and H~bond energies ($-\eta h_{ij}$), which are
2-body terms, are split among $6$ tetrahedra, whence the
$1/6$~prefactor in Eq.~\eqref{eq:ham}. On the contrary, the 3-body
weakening terms ($\eta c_\saf{x} h_{ij} \nocc{x}{k} /6$) are split
between $2$ tetrahedra, thus the $1/6$ factor is absorbed in the
prefactor, while a $1/2$ factor is left in the tetrahedron
hamiltonian.

\section{First order approximation}

We perform the investigation by means of a generalized first order
approximation on a tetrahedral cluster. In the previous
paper~\cite{PrettiBuzano2004}, we have introduced the
approximation in the variational approach~\cite{An1988}, as a
particular choice of the largest clusters left in the entropy
expansion (basic clusters). Such a choice, sometimes denoted as
cluster-site approximation~\cite{Oates1999} (the only clusters to
be taken into account in the expansion are basic clusters and
single sites), has not only the advantage of high simplicity, but
also of relative accuracy, which has been recognized for different
models~\cite{Oates1999,BuzanoDestefanisPelizzolaPretti2004}. The
basic clusters are a number of irregular tetrahedra, namely, 4 out
of 24 tetrahedra sharing a given site, as sketched in
Fig.~\ref{fig:cactustetraedro}. This choice turns out to coincide
with the (generalized) first order approximation (on a
tetrahedron), also equivalent to the exact calculation on a Husimi
lattice~\cite{Pretti2003}, whose (tetrahedral) building blocks are
just arranged as in Fig.~\ref{fig:cactustetraedro}(b). In the
present paper, we employ the latter approach, which, not having to
treat cluster probability distributions explicitly, is numerically
more convenient.

Husimi lattice thermodynamics can be studied exactly, in a
numerical way, by solving a suitable recursion
relation~\cite{Pretti2003}, since the system is locally treelike.
First of all, we have to choose, as a Husimi lattice hamiltonian,
the following expression
\begin{equation}
  \ham' =
  \sum_{\langle \alpha,\beta,\gamma,\delta \rangle}
  \ham_{i_\alpha i_\beta i_\gamma i_\delta}
  ,
  \label{eq:hamprime}
\end{equation}
obtained from Eq.~\eqref{eq:ham} by understanding that the sum
runs over tetrahedra in the treelike system only, and removing the
$1/6$ prefactor. Let us notice that the latter change is required,
in order to obtain equal internal energy densities (internal
energies per site) for the Husimi lattice and the ordinary lattice
system. If we denote the tetrahedron configuration probability by
$p_{ijkl}$, and assume that it is equal on every tetrahedron, the
internal energy density can be written as
\begin{equation}
  u = \sum_{i=0}^4 w_i \sum_{j=0}^4 w_j \sum_{k=0}^4 w_k
  \sum_{l=0}^4 w_l p_{ijkl} \ham_{ijkl}
  ,
  \label{eq:intenergy}
\end{equation}
where $w_i$ is the multiplicity of the $i$-th site configuration,
equal to $w$ for the nonbonding configuration of water ($i=3$) and
equal to $1$ otherwise (see Tab.~\ref{tab:configurazioni}). One
then has to define partial partition functions~(PPFs) in the
following way. Let us consider a single branch of a Husimi tree,
made up of tetrahedral blocks, and a corresponding partial
hamiltonian, obtained by Eq.~\eqref{eq:hamprime} with the sum
restricted to tetrahedra in the branch. The PPF~$Q_i$ is defined
as a sum of the Boltzmann weights of the partial hamiltonian over
the configurations of the branch minus the base site (that is why
the PPF depends on a site configuration variable $i$). Working in
the grand-canonical ensemble, as we are interested in, we also
have to take into account chemical potential contributions, which
is done by replacing the tetrahedron hamiltonian~$\ham_{ijkl}$
with
\begin{equation}
  \tilde{\ham}_{ijkl} = \ham_{ijkl}
  - \mu_\saf{x} \frac{\nocc{x}{i} + \nocc{x}{j} + \nocc{x}{k} + \nocc{x}{l}}{4}
  \label{eq:tetrahamtilde}
  ,
\end{equation}
where the usual convention on repeated indices holds. Let us
notice that of course the PPF tends to infinity in the
thermodynamic limit, that is for an ``infinite generation
branch'', therefore it is convenient to define a normalized PPF
$q_i \propto Q_i$, for instance in such a way that
\begin{equation}
  \sum_{i=0}^4 w_i q_i = 1
  .
  \label{eq:norm}
\end{equation}
In this way, $q_i$ represents the $i$-th configuration probability
that the base site would have if it were not attached to any other
branch.

Let us now consider again the single branch of our Husimi tree. In
the infinite generation limit, and in the hypothesis of a
homogeneous system, the subbranches attached to the first
tetrahedral block should be equivalent to the main one, so that
one can write the recursion relation
\begin{equation}
  q_i = y^{-1}
  \sum_{j=0}^4 w_j
  \sum_{k=0}^4 w_k
  \sum_{l=0}^4 w_l
  e^{-\tilde{\ham}_{ijkl}/T}
  \left( q_j q_k q_l \right)^3
  ,
  \label{eq:rec}
\end{equation}
where the sums run over configuration variables in the tetrahedron
except~$i$, $T$ is the temperature expressed in energy units
(whence entropy will be expressed in natural units), and $y$~is a
normalization constant, imposed by Eq.~(\ref{eq:norm}). The
recursion relation can be iterated numerically to determine a
fixed point, representing the PPF of a branch whose base site lies
in the bulk of the Husimi tree (generally denoted as Husimi
lattice). Husimi lattice properties are equivalent to those
obtained by the cluster-site approximation~\cite{Pretti2003}. We
can compute the site probability distribution~$p_i$, by
considering the operation of attaching $4$ equivalent branches to
the given site. We obtain
\begin{equation}
  p_i = z^{-1} q_i^4  ,
  \label{eq:pjoint}
\end{equation}
where
\begin{equation}
  z = \sum_{i=0}^4 w_i q_i^4
  \label{eq:zjoint}
\end{equation}
provides normalization. We can also compute the tetrahedron
probability distribution, by considering the operation of
attaching $3$ equivalent branches to each site of a given
tetrahedron, yielding
\begin{equation}
  p_{ijkl} \propto
  e^{-\tilde{\ham}_{ijkl}/T}
  \left( q_i q_j q_k q_l \right)^3
  ,
\end{equation}
where of course the proportionality constant is determined by
normalization. From the knowledge of the tetrahedron probability
distribution $\{p_{ijkl}\}$ one can compute the thermal average of
every observable in the first order approximation, the internal
energy density by Eq.~\eqref{eq:intenergy}, and the
grand-canonical free energy by Eq.~(10) in
Ref.~\onlinecite{PrettiBuzano2004}. According to Eq.~(31) in
Ref.~\onlinecite{Pretti2003}, the latter can be also related to
normalization constants as
\begin{equation}
  \omega = - T \left( \ln y - 2 \ln z \right)
  ,
\end{equation}
where $y$~is the normalization constant of the recursion
relation~(\ref{eq:rec}) and $z$~is given by Eq.~(\ref{eq:zjoint}).
Finally, the entropy density can be computed as
\begin{equation}
  s = \frac{u - \mu_\saf{x} \rho_\saf{x} - \omega}{T}
  ,
\end{equation}
where $u-\mu_\saf{x}\rho_\saf{x}$ has formally the same expression
as~$u$ in Eq.~\eqref{eq:intenergy}, with the tetrahedron
hamiltonian~$\ham_{ijkl}$ replaced by~$\tilde{\ham}_{ijkl}$.

\section{Results}

In order to investigate the model properties, we fix a set of
parameters, as a result of several attempts. First of all, we take
$\epsilon_\saf{ww}/\eta = 0.3$. This value of water-water Van der
Waals interaction is equal to the one employed for the very
detailed analysis by Roberts et
al.~\cite{RobertsKarayiannakisDebenedetti1998}, though a bit
larger than previously employed by us~\cite{PrettiBuzano2004}.
Anyway, this choice accounts for the greater binding energy of
hydrogen bonds with respect to Van der Waals interactions. As far
as the multiplicity of nonbonding water configurations is
concerned, we set $w = 20$ (as in the previous work), which mimics
the high directionality of hydrogen bonds. For neat water, it is
necessary to set this parameter large enough to let anomalous
properties appear, but further increase does not change
qualitatively the phase diagram and the thermodynamic properties.
As far as the weakening parameters are concerned, for water we
choose $c_\saf{w}=0.5$, which, given the other parameter values,
corresponds to a situation without a reentrant spinodal, in
agreement with most recent molecular dynamics
results~\cite{Stanley2003}.

\subsection{Pure water properties}

\input{tp.tex}

Since the parameter choice is slightly different from the former
work~\cite{PrettiBuzano2004}, we first reconsider neat water
properties, taking the limit $\mu_\saf{s} \to -\infty$. This
analysis is mainly meant to show that the model behavior still
remains consistent with a ``realistic'' phase diagram, that is,
with several experimental evidences as well as simulation
predictions. In fact, we have observed that relatively small
variations in parameter values may give rise to quite dramatic
changes in the phase behavior~\cite{PrettiBuzano2004}, as observed
also in other models of
water~\cite{TruskettDebenedettiSastryTorquato1999}. The
temperature-pressure phase diagram is reported in
Fig.~\ref{fig:tp}. Imposing homogeneity, our analysis includes
both thermodynamically stable and metastable (supercooled) phases,
even if stability is not investigated. Let us notice that pressure
can be determined as $P = -\omega$, where $\omega$ is the
grand-canonical potential per site, introduced in the previous
section. We have assumed the volume per site equal to unit, so
that pressure is expressed in energy units. The appropriate order
parameter is the density, i.e., the probability~$\rho_\saf{w}$
that a site is occupied by a water molecule, which can be
evaluated from the formula
\begin{equation}
  \rho_\saf{x} = \sum_{i=0}^4 w_i p_i \nocc{x}{i}
  .
  \label{eq:density}
\end{equation}

We find two different first order transition lines, terminating in
two different critical points. The positively sloped line, at
lower pressures, corresponds to the coexistence of a very low
density phase and a high density phase, and represents the
ordinary vapor-liquid transition. The other one, negatively sloped
and placed at higher pressures, corresponds to coexistence between
the high density liquid and and a lower density one. The related
critical point may reasonably represent the so-called second
critical point, which has been conjectured and observed in
simulations, and of which also some experimental evidences have
been given. As one could expect, the low density liquid turns out
to be more hydrogen bonded than the high density one. We find a
density maximum as a function of temperature for liquid coexisting
with vapor (and at constant pressure as well), and the temperature
of maximum density slightly decreases upon increasing pressure, as
observed in experiments.

We also report the liquid phase spinodals and the Kauzmann line.
Details about the (semi-analytical) calculation of spinodals,
which allows to determine density response functions as well, are
given in the previous paper~\cite{PrettiBuzano2004}. The limit of
stability of the liquid phase (spinodal) is the locus at which the
metastable liquid ceases to be a minimum of (a variational form
of) the free energy, and becomes a saddle point. On the contrary,
the Kauzmann line is the locus at which the liquid phase entropy
vanishes, and corresponds to the ideal glass transition. It can be
easily determined numerically. As previously mentioned, we do not
observe a reentrance of the liquid-vapor spinodal in the positive
pressure half-plane, which is actually a possibility of our model,
for a different parameter choice, namely, for higher values of the
weakening parameter. The reentrant spinodal scenario was one of
the conjectures invoked to explain thermodynamic anomalies of
liquid water, and in particular of the divergent-like behavior of
response functions in the supercooled
regime~\cite{Speedy1982I,ZhengDurbenWolfAngell1991}, even if at
the moment the second critical point scenario is believed to be
more realistic~\cite{Stanley2003}.

For the present parameter choice, in our model the critical point
lies just above the Kauzmann line, while most of the high-low
density liquid transition lies in the negative entropy region.
This is consistent with the experimental observation of two
different forms of amorfous ice in this region, while the
``underlying'' liquid phase is just an extrapolation of the
equation of state for the liquid. The Kauzmann line displays a
cusp (actually a slight discontinuity), while intersecting the
high-low density liquid transition. It is noticeable that a
similar feature has been predicted also by a recent analysis of
the potential energy landscape of simulated water, performed by
Sciortino and coworkers~\cite{SciortinoLaNaveTartaglia2003}, on
the basis of the inherent structure theory.

\input{pcost.tex}

Let us also report the density response functions and the specific
heat of the liquid at constant pressure $P/\eta=0.015,0.030$,
roughly corresponding to $1/5,2/5$ of the liquid-vapor critical
pressure. Also for these calculations, details are reported in
Ref.~\onlinecite{PrettiBuzano2004}. We find anomalous behavior,
qualitatively similar to that of real liquid water. The first
response function we consider is the thermal expansion coefficient
$\alpha_P = (-\partial\ln\rho/\partial T)_P$, which, from
statistical mechanics, is known to be proportional to the
entropy-volume cross-correlation. For ordinary fluids,
$\alpha_P$~is always positive, i.e., the local entropy and the
local specific volume are positively correlated. On the contrary,
for our model $\alpha_P$ (Fig.~\ref{fig:pcost}, top panel) is
anomalous. As temperature is lowered, the expansion coefficient
vanishes (at the temperature of maximum density), and then becomes
negative. Of course, we do not observe a really divergent behavior
of this coefficient, but a pronounced peak instead. Let us notice
that, upon increasing pressure, the peak is observed to become
broader, indicating that the liquid is becoming more normal, in
agreement with experiments. The trend of the isothermal
compressibility $\kappa_T = (\partial\ln\rho/\partial P)_T$ is
also anomalous (Fig.~\ref{fig:pcost}, middle panel). For a typical
liquid, $\kappa_T$ decreases as one lowers temperature, because it
is proportional to density fluctuations, whose magnitude decreases
upon decreasing temperature. On the contrary, we observe that
$\kappa_T$, once reached a minimum, begins to increase upon
decreasing temperature. The constant pressure specific heat $c_P =
(T\partial s/\partial T)_P$ (Fig.~\ref{fig:pcost}, bottom panel)
displays a completely analogous behavior, with the minimum
occurring at a higher temperature.

\subsection{Solution properties}

Let us now consider an ideal inert solute molecule, with no Van
der Waals interaction with water ($\epsilon_\saf{ws}=0$), nor with
other solute molecules ($\epsilon_\saf{ss}=0$), and no weakening
effect on H bonds ($c_\saf{s}=0$). As a first analysis of the
model mixture of water with this kind of (hydrophobic) solute, let
us investigate phase diagrams at constant temperature and constant
pressure, corresponding to the ``cuts'' reported in
Fig.~\ref{fig:tp}. We take into account, as a composition
variable, the solute molar fraction, defined as
\begin{equation}
  x_\saf{s} = \frac{\rho_\saf{s}}{\rho_\saf{w} + \rho_\saf{s}}
  ,
\end{equation}
where $\rho_\saf{w}$ and $\rho_\saf{s}$ are determined by
Eq.~\eqref{eq:density}. In Fig.~\ref{fig:xptx} (top panel), we
report a constant temperature phase diagram, computed at
$T/\eta=0.34$. We can observe a first order transition between a
water-rich and a solute-rich phase, which arises continuously from
the vapor-liquid transition of neat water, as solute concentration
is increased. To determine this transition, we have fixed several
different values of water chemical potential $\mu_\saf{w}$, then
we have determined numerically the value of solute chemical
potential $\mu_\saf{s}$, for which both phases had the same
pressure $P=-\omega$. The phase-separated (coexistence) region
occupies large part of the diagram, that is, the molar fraction of
solute which can be dissolved into water, without giving rise to
phase separation, is very small. The behavior is very similar to
that of an ordinary solution of two disaffine chemical species (at
a temperature higher than the critical one for the solute), with
no peculiar anomaly related to the hydrophobic effect. Notice that
actually we cannot observe any phase transition for neat solute
($x_\saf{s}=1$), because we have described it as a perfect gas,
with $\epsilon_\saf{ss}=0$. We have also computed a constant
pressure phase diagram at $P/\eta=0.015$, which we have reported
in Fig.~\ref{fig:xptx} (bottom panel). Here, to compute the
transition, we have adjusted numerically both chemical potentials
$\mu_\saf{w}$ and $\mu_\saf{s}$ to impose equality between
pressures of both phases and the reference one. As expected, also
in this case we observe no phase transition for pure solute,
whereas a clearly anomalous behavior is observed for the
water-rich phase boundary. Upon decreasing temperature, near
$T/\eta \approx 0.31$, the slope of this curve begins to change
rapidly. Such a behavior gives rise to an absorption coefficient
$x_\saf{s}^{w}/x_\saf{s}^{s}$ (the superscripts denoting the
water-rich and the solute-rich phase, respectively) with a minimum
around $T/\eta \approx 0.34$, well above the temperature of
maximum density at that pressure (see Fig.~\ref{fig:pcost}). This
is typical signature of hydrophobic
effect~\cite{WidomBhimalapuramKoga2003}.
\input{xptx.tex}
\input{tdxs.tex}

Let us now turn to the transfer properties of a single molecule in
water, that is, to a dilute solution. Large amounts of
experimental data are available for this case~\cite{BenNaim1987}.
According to the Ben-Naim standard~\cite{BenNaim1987}, a transfer
process (for a chemical species $\saf{x}$) can be characterized by
means of the pseudo-chemical potential $\mu^{*}_\saf{x}$ , related
to the ordinary chemical potential $\mu_\saf{x}$ by the formula
\begin{equation}
  \mu_\saf{x} = \mu^{*}_\saf{x} + T\log\rho_\saf{x}
  .
\end{equation}
The use of pseudo-chemical potentials is meant to remove
translational entropy contributions, which are not directly
related to the solvation process. The solvation free energy per
molecule $\Delta g^*_\saf{x}$, that is, the free energy of
transfer for a molecule $\saf{x}$ from the gas phase to the liquid
phase, can then be defined as
\begin{equation}
  \Delta g^*_\saf{x} = \mu^{*l}_\saf{x} - \mu^{*g}_\saf{x}
  ,
  \label{eq:mudg}
\end{equation}
where $\mu^{*l}_\saf{x}$ and $\mu^{*g}_\saf{x}$ denote
pseudo-chemical potentials in the liquid and gas phase,
respectively. If the gas and liquid phases coexist in equilibrium,
the ordinary chemical potentials for the given species must be
equal, so that we obtain
\begin{equation}
  \Delta g^*_\saf{x} = -T \ln
  \frac{\rho_\saf{x}^l}{\rho_\saf{x}^g}
, \label{eq:realdg}
\end{equation}
where $\rho_\saf{x}^l$ and $\rho_\saf{x}^g$ denote densities for
the $\saf{x}$ species in the two phases. Derived quantities, of
interest in experiments, are the solvation entropy
\begin{equation}
  \Delta s^*_\saf{x} = - \frac{\partial \Delta g^*_\saf{x}}
  {\partial T}\biggl\lvert_P
  ,
  \label{eq:constpentropy}
\end{equation}
the solvation enthalpy
\begin{equation}
  \Delta h^*_\saf{x} = \Delta g^*_\saf{x} + T \Delta s^*_\saf{x}
  ,
\end{equation}
and the solvation heat-capacity
\begin{equation}
  \Delta {c_P}^*_\saf{x} = \frac{\partial\Delta h^*_\saf{x}}
  {\partial T}\biggl\lvert_P
  .
\end{equation}
In principle, we should distinguish between derivatives taken at
constant pressure (as stated by definition) or along the
liquid-vapor equilibrium curve. In particular, we could not even
use Eq.~\eqref{eq:realdg}, because we would move out of the
equilibrium curve. Nevertheless, we have verified that the
difference between the two sets of results is negligible, in
agreement with experimental observations~\cite{BenNaim1987}, and
one can usually take the ``equilibrium'' derivative without
further care.

Let us start studying solvation properties for the ideal inert
solute, in the framework of our model. Water parameters are fixed
as in the previous case. The temperature trends of the free
energy, entropy, and enthalpy of transfer are given in
Fig.~\ref{fig:tdxs}a; the transfer heat capacity in
Fig.~\ref{fig:tdxs}c. In order to compare with experimental
data~\cite{CrovettoFernandez-PriniJapas1982,BenNaim1987}, also
reported in Fig.~\ref{fig:tdxs}b and~\ref{fig:tdxs}d,
respectively, all quantities are evaluated at liquid-vapor
coexistence, and for very low solute density with respect to water
density (dilute solution limit). In practical calculations, we
have adjusted numerically water and solute chemical potentials, in
order to impose the equilibrium condition (equal pressure) between
liquid and vapor, and to fix the solute molar fraction.
Nevertheless, we have also verified that, only for the perfectly
inert solute, concentration does not affect the results at all, so
that we could also set an arbitrary value for the solute chemical
potential. As shown in the previous section, our parameter set for
pure water corresponds to a liquid-vapor critical temperature
$T/\eta \approx 0.52$, and to a temperature of maximum density for
the liquid phase around $T/\eta \approx 0.32$, at low pressure.
Therefore, in order to represent roughly the experimental
temperature range (between $0^\circ\,\mathrm{C}$ and
$300^\circ\,\mathrm{C}$) we report model results between $T/\eta =
0.31$ (just below the temperature of maximum density for pure
liquid water) and $T/\eta = 0.40$ (about half way between the
previous temperature and the critical temperature). Remarkably, it
turns out that the model displays the defining features of
hydrophobic solvation. The solvation free energy is positive and
large, while the solvation entropy and enthalpy are negative at
low temperatures and become positive upon increasing temperature.
The solvation heat capacity is positive and large, and also the
decreasing trend with temperature is basically reproduced. A
slightly increasing trend at high temperature is related to the
the fact that we are approaching the liquid-vapor critical point.
Negative solvation entropy at low (room) temperature is a clear
indication that solute insertion into the mixture orders the
system. The corresponding positive (unfavorable) contribution to
free energy compensates a negative (favorable) enthalpic
contribution, giving rise to a positive solvation free energy. At
higher temperature, enthalpic and entropic contributions change
sign, but they still have the same compensating trend. The model
also predicts two different temperatures at which the transfer
enthalpy and entropy vanish (see Fig.~\ref{fig:tdxs}a), as
observed in experiments (Fig.~\ref{fig:tdxs}b).
\input{tdxw.tex}
The whole observed behavior is to be ascribed to the
thermodynamics of H bonding and, in order to rationalize this fact
in the model framework, we have also analysed transfer quantities,
upon removing H bond interactions (see
Fig.~\ref{fig:tdxs}a,\ref{fig:tdxs}c). As reasonable, the results
are quite similar in the high temperature regime, where there is a
high probability that H bonds are broken by thermal fluctuations,
whereas they change more and more dramatically upon decreasing
temperature, and, in particular, the regions of negative transfer
entropy and enthalpy completely disappear. This facts confirm that
H bonding is the key element for system ordering, upon insertion
of an inert molecule. Accordingly, also the increasing trend of
the heat capacity upon decreasing temperature is suppressed. The
process is now dominated by enthalpy, with a large and positive
transfer free energy (but without a maximum), and a positive
transfer entropy. Transfer quantities now behave qualitatively as
observed in solvation experiments of noble gas molecules in
ordinary liquids~\cite{DaviesDuncan1967,BenNaim1987}, and are
relatively independent of temperature. Actually, a water molecule
for which H bond formation has been ``turned off'', can be viewed
as a nonpolar molecule with only Van der Waals interaction
energy~$\epsilon_{\saf{ww}}$.

Let us now consider also the solvation of water in its own pure
liquid. Corresponding transfer quantities obtained by the model
are displayed in Fig.~\ref{fig:tdxw}a, where we have reduced the
temperature interval, in order to compare with available
experimental results~\cite{BenNaim1987}, reported in
Fig.~\ref{fig:tdxw}b. With respect to the inert molecule case,
here absolute values of solvation free energy and entropy are
considerably smaller. Enthalpy, rather than entropy, dominates the
solvation process, while all quantities are relatively independent
of temperature. These features characterize a regular transfer
process, like the solvation of an ordinary fluid molecule from a
gas phase into its pure liquid phase. In this case, upon removing
H bond interactions (thin lines in Fig.~\ref{fig:tdxw}a), very
little changes are observed. Let us discuss two issues about these
results. First, the fact that so little changes are caused by
turning on or off H bonds can be rationalized on the basis of the
microscopic model interactions. Insertion of a water molecule into
pure liquid water should imply in principle the formation of new H
bonds, but the model is such that insertion of a new water
molecule also weakens other H bonds in its neighborhood, and the
two effects nearly compensate each other. Second, let us notice
that solvation enthalpy decreases upon increasing temperature,
that is, the solvation heat capacity is negative, in contrast with
experiments. We do not have an explanation for this fact, but we
can observe that it is basically unchanged when H bonds are turned
off, that is, when the model is reduced to describe a ``regular''
solvation process. This suggest that there is probably a
limitation of the lattice description, that anyway has nothing to
do with peculiarities of water. Indeed, the effect is
quantitatively small, so that it is hidden by other large
(enthalpic and entropic) effects, observed in the case of
hydrophobic solvation.

\subsection{``Nonideal'' solute and entropy convergence}

\input{entconv.tex}
So far, we have always turned off all interactions involving
solute molecules, except excluded volume. Here we report some
results concerning the role of nonzero solute-water interaction
parameters ($\epsilon_\saf{ws}$, $c_\saf{s}$), still assuming that
solute molecules have no relevant interaction with one another
($\epsilon_\saf{ss} = 0$). In order to have a single parameter to
be varied, we have performed our investigation for increasing
values of the solute weakening parameter $c_\saf{s}$, and defined
solute-water interaction $\epsilon_\saf{ws}$ according to the
following proportionality condition
\begin{equation}
  \epsilon_\saf{ws}/\epsilon_\saf{ww} = c_\saf{s}/c_\saf{w}
  .
  \label{eq:proporz}
\end{equation}
At the beginning, we took this assumption as a simple trial, but
the results we obtained were quite interesting, so that we have
carried on with the analysis. As far as transfer free energies are
concerned, we have observed a qualitatively unchanged behavior,
with a broad maximum at some temperature, and free energy values
getting smaller and smaller, upon increasing $\epsilon_\saf{ws}$
and $c_\saf{s}$. On the contrary, we have observed peculiar
features concerning entropies, actually related to one another,
which are displayed in Fig.~\ref{fig:entconv}a. Still upon
increasing $\epsilon_\saf{ws}$ and $c_\saf{s}$, the temperature of
zero entropy is progressively shifted towards higher values, while
different entropy curves converge in a very narrow temperature
range, relatively close to zero entropy temperatures. Moreover,
entropy values at convergence are negative and relatively small.
As one can observe in Fig.~\ref{fig:entconv}b, all of these
features correspond remarkably well to phenomenology observed for
the series of noble gases~\cite{BenNaim1987}, in particular the
entropy convergence, which has attracted some interest, due to the
fact that a similar effect has been observed for the entropies of
protein unfolding~\cite{Lee1991pnas}. Because of this unexpected
result, we have tried and justified a posteriori the working
hypothesis~\eqref{eq:proporz}, and actually a naive explanation
may be the following one. As a first approximation, different
hydrophobic species, such as noble gases, may be viewed as hard
spheres distinguished by their diameter, that is their volume,
only. In this way, the proportionality condition can be conceived
as a trick that, in the framework of the lattice model, mimics the
fact that ``a fraction'' $c_\saf{s}/c_\saf{w}$ of the site
occupied by the solute actually ``behaves like water''. As a
consequence, higher values of the ratio $c_\saf{s}/c_\saf{w}$
should correspond to smaller solute molecules, which turns out to
be consistent, comparing Figs.~\ref{fig:entconv}a
and~\ref{fig:entconv}b. Let us notice also that, for parameter
values as in Fig.~\ref{fig:entconv}a, the ``fractions that behaves
like solute'' ($1-c_\saf{s}/c_\saf{w}$) correlate well with the
squares of the hard sphere diameters of the corresponding
substances~\cite{Garde1996}, which is consistent with the common
assumption that hydrophobicity is proportional to exposed surface.

\section{Discussion and conclusions}

In this paper we have considered a 3-dimensional lattice fluid
model of water, which we had previously shown to exhibit realistic
thermodynamic anomalies~\cite{PrettiBuzano2004}, and extended the
model to describe aqueous solutions. The motivation for this work
resides mainly in the interest for the hydrophobic effect, whose
relevance for biological processes such as protein folding, taking
place in aqueous solutions, has been more and more recognized in
the latest years. Moreover, a simplified but accurate modelling of
water is an appealing issue, in view of investigations on such
processes, because detailed water models may be extremely time
consuming~\cite{SouthallDillHaymet2002}.

As far as water is concerned, our model is a simplified version of
a previous model proposed by Roberts and
Debenedetti~\cite{RobertsDebenedetti1996}, without a distinction
between hydrogen bond donors and acceptors. In the framework of
this model, the microscopic description of water anomalies, is
essentially based on the competition between an isotropic (Van der
Waals like) interaction and an highly directional (H~bonding)
interaction, and on the difference between the respective optimal
interaction distances. In the lattice environment, the latter is
taken into account by a trick, that is, the weakening effect of a
water molecule near a formed bond. The same assumptions also
accounts for possible perturbations of the electronic density, due
to interaction with other water molecules. We have extended the
weakening trick to take into account the presence of a different
chemical species (solute), with no internal degrees of freedom
(bonding arms). Calculations have been performed in a generalized
first-order approximation on a tetrahedral cluster, which requires
small computational effort, and had been shown to be quite
accurate for the pure water model~\cite{PrettiBuzano2004}.

Due to the fact that we have chosen slightly different interaction
parameters for the present investigation, we have first analysed
again pure water behavior. At constant pressure, the typical
thermodynamic anomalies are reproduced, with a density maximum,
and a minimum of isothermal compressibility and specific heat. In
the ordinary temperature and pressure region, the temperature of
maximum density decreases upon increasing pressure, as observed in
experiments. Moreover, as far as the supercooled regime is
concerned, there is still evidence of a second (metastable)
critical point, which terminates a line of coexistence between two
liquid phases at different densities. Let us recall that the pure
water model could predict, for different values of the weakening
parameter $c_\saf{w}$, two different scenarios, that is, with or
without a reentrant spinodal. The present parameter choice
predicts a nonreentrant spinodal. The reentrant spinodal scenario
was the first conjecture put forth to justify water anomalies,
whereas the most recent and accurate molecular dynamics
simulations of water suggest a scenario with a nonreentrant
spinodal and a metastable liquid-liquid critical
point~\cite{Stanley2003}. As far as the critical point is
concerned, in our calculation, it lies at some temperature just
above the Kauzmann line, at which the configurational entropy
vanishes, while the Kauzmann line displays a cusp upon crossing
the metastable coexistence line. All of these features turn out to
be in a remarkably good agreement with the inherent state analysis
of the potential energy landscape of simulated water, recently
performed by Sciortino and
coworkers~\cite{SciortinoLaNaveTartaglia2003}.

As far as the solution model is concerned, we have mainly
considered solutions of inert, that is, hydrophobic solutes. First
of all, we have investigated phase diagrams for arbitrary solute
concentration, pointing out a minimum of solubility as a function
of temperature, as typical for hydrophobic solutions.
We have investigated in more detail the dilute solution limit, at
liquid-vapor equilibrium, for which many experimental data are
available. Solvation quantities turn out to exhibit peculiar
features that are believed to be the fingerprints of
hydrophobicity. The solvation free energy is positive (unfavorable
solvation), while entropy and enthalpy are negative at low
temperatures and positive at high temperatures. The solvation heat
capacity is large and decreases upon increasing temperature. These
results compare qualitatively well with solvation experiments for
noble gases in water. Let us notice that a previous lattice model
by Besseling and Lyklema~\cite{BesselingLyklema1997}, based on a
different description of water interactions, was also able to
account for the qualitative behavior of free energy, enthalphy,
and entropy of transfer. Nevertheless, it failed in reproducing
the correct temperature trend of the transfer heat capacity,
which, according to some authors~\cite{SouthallDillHaymet2002} is
a key feature, revealing the cooperative nature of the hydrophobic
effect. Let us recall, by the way, that the same difficulty about
heat capacity is encountered by the information theory
approach~\cite{ArthurHaymet1999}.

We have investigated explicitly on the effect of H bonding, in the
framework of our model, performing calculations also when this
interaction is completely turned off. In this case, we have
obtained transfer quantities that approach the ones computed {\em
with} H bonds at high temperatures, but that largely deviates from
them upon decreasing temperature, that is, in the region were H
bonding begins to dominate. In particular, we have observed that,
while disaffinity between solute and solvent is left (the
solvation free energy is still positive), this is mainly of
enthalpic nature. Both the enthalpy and entropy of solvation
remain positive at all temperatures, so that also the typical
strong temperature dependence of hydrophobic solvation,
disappears.

We have also taken into account the solvation process of water
into its own pure liquid, for which experimental data are
available. We have found qualitative agreement, as far as the
values of transfer free energy, entropy and enthalpy are
concerned, but we have observed some discrepancy in the
temperature dependence of enthalpy, indicating a negative
solvation heat capacity, in disagreement with experiments. We have
verified that the same kind of discrepancy can be observed if H
bonding is turned off, that is, for an ordinary lattice gas.
Therefore, we suggest that the discrepancy is to be related to an
intrinsic limitation of the lattice environment, that has nothing
to do with peculiarities of the water model. The effect is
relatively small, so that it is completely invisible, when the
dominant effect of H bonds is introduced.

Let us notice that the results concerning transfer quantities are
qualitatively similar to those observed for a 2-dimensional
lattice model recently investigated by
us~\cite{BuzanoDestefanisPretti2005}, which in turn is a
simplified ``lattice version'' of the Mercedes Benz model
investigated by Dill and coworkers. In spite of similarities in
the experimental temperature range, there is one major difference
between the two models. The 2-dimensional model is not able to
account for the metastable critical point of water, probably due
to impossibility of reproducing a high density ordered packing of
water molecules. As a consequence, thermodynamic anomalies are
entirely due to the presence of a reentrant spinodal, which is no
longer believed to be a realistic scenario for real
water~\cite{Stanley2003}. In this sense, the present 3-dimensional
model seems to provide a more coherent view of water
thermodynamics, relating each other neat water anomalies and
hydrophobic effect.

Moreover, we have shown that, with quite a reasonable assumption
for the solute interaction parameters
$\epsilon_\saf{ws}$,$c_\saf{s}$ (attempting to describe solutes of
different volume in the lattice framework), the model is able to
reproduce also the entropy convergence phenomenon, in a
qualitatively correct way. Such a phenomenon has been
theoretically described, for instance, by the simplified molecular
theory of Debenedetti and
coworkers~\cite{AshbaughTruskettDebenedetti2002}. Moreover, an
almost quantitative explanation has been proposed by Pratt and
coworkers~\cite{Garde1996}, on the basis of the information theory
approach, which we mentioned in the Introduction. In the cited
work, the authors argue that entropy convergence, and in
particular the negative entropy at convergence, are related to the
weak temperature dependence of free volume fluctuations in liquid
water, that is, isothermal compressibility. Although we cannot
provide a clear explanation of why our lattice model, with the
proportionality assumption~\eqref{eq:proporz}, exhibits
qualitatively correct behavior, let us notice that such result is
somehow consistent with the previous explanation. In fact, entropy
convergence occurs very close to the minimum of isothermal
compressibility, as one can argue from Fig.~\ref{fig:pcost}, that
is, in a region where compressibility is nearly constant, as a
function of temperature. Nonetheless, the proportionality
assumption~\eqref{eq:proporz} still remains not well justified.

Let us finally recall that our model, at least in the present
treatment, is not able to provide microscopic structural details
as simulations do, but its most appealing feature is simplicity.
In the present paper we have shown that, in spite of this, the
model yields a qualitatively coherent description of peculiar
thermodynamics of water, not only as a pure substance but also as
a solvent, and is consistent with predictions based on much more
sophisticated models and simulations. Moreover, thanks to the
3-dimensional embedding, it may be suitable for quite a realistic
analysis of more complex, for instance polymeric, solutes. We are
going to report about such investigations in a forthcoming
article.


\input{main.bbl}
\end{document}

%% file: reticolo.tex
\begin{figure}[t!]
  \includegraphics*[20mm,193mm][100mm,268mm]{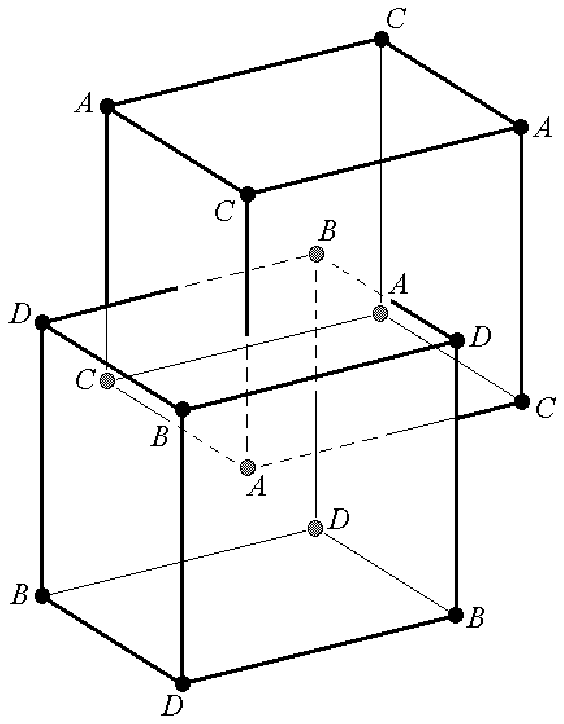}
  \caption
  {
    Two conventional cells of the body-centered cubic lattice:
    $A,B,C,D$ denote 4 interpenetrating face-centered cubic
    sublattices.
  }
  \label{fig:reticolo}
\end{figure}

%% file: molecole.tex
\begin{figure}[t!]
  \includegraphics*[20mm,193mm][100mm,268mm]{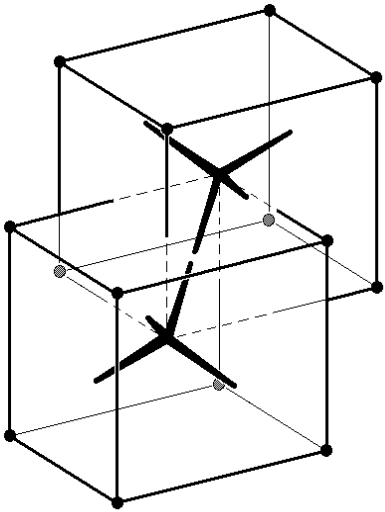}
  \caption
  {
    Two model molecules forming a H~bond.
    The lower molecule is in the $i=1$ configuration,
    the upper one is in the $i=2$ configuration.
  }
  \label{fig:molecole}
\end{figure}

%% file: cactustetraedro.tex
\begin{figure}[t!]
  \resizebox{80mm}{!}{\includegraphics*[20mm,195mm][90mm,277mm]{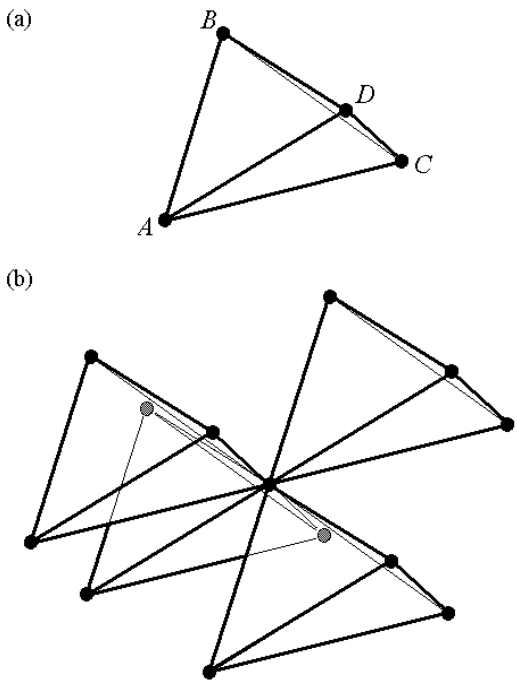}}
  \caption
  {
    (a)~Basic cluster (irregular tetrahedron):
    $A,B,C,D$ denote sites in the 4 corresponding sublattices. $AB$,
    $BC$, $CD$, and $DA$ are NN pairs; $AC$ and $BD$ are second neighbor
    pairs.
    (b)~Husimi tree structure corresponding to the generalized first
    order approximation on the tetrahedron.
  }
  \label{fig:cactustetraedro}
\end{figure}

%% file: tp.tex
\begin{figure}[t!]
  \includegraphics*[40mm,195mm][114mm,250mm]{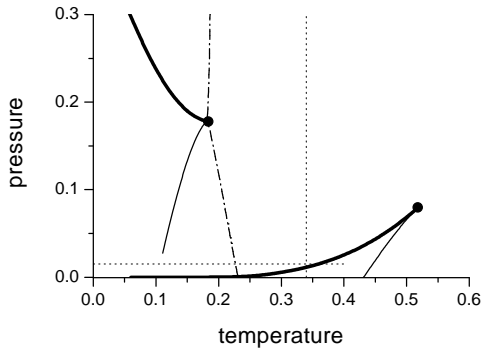}
  \caption
  {
    Pressure ($P/\eta$) vs temperature ($T/\eta$) phase diagram
    for pure water with $\epsilon_\saf{ww}/\eta=0.3$, $c_\saf{w}=0.5$, and $w=20$.
    Thick solid lines denote (first order) phase transitions;
    circles are critical points.
    Thin (solid and dash-dotted) lines denote
    the spinodals and the Kauzmann line, respectively.
    Dotted straight lines denote constant temperature
    ($T/\eta=0.34$) and constant pressure ($P/\eta=0.015$)
    ``cuts'', corresponding to the solution phase diagrams,
    reported in Fig.~\ref{fig:xptx}.
  }
  \label{fig:tp}
\end{figure}

%% file: pcost.tex
\begin{figure}[t!]
  \includegraphics*[40mm,115mm][114mm,250mm]{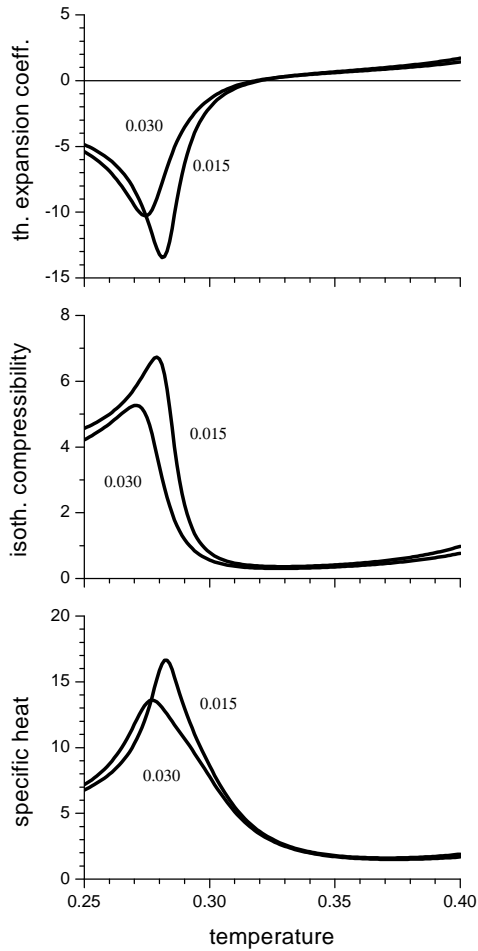}
  \caption
  {
    Response functions at constant pressure ($P/\eta=0.015,0.030$)
    for pure liquid water as a function of temperature ($T/\eta$),
    for $\epsilon_\saf{ww}/\eta=0.3$, $w=20$, and $c_\saf{w}=0.5$.
    From top to bottom, we show
    the isobaric thermal expansion coefficient ($\eta \alpha_P$),
    the isothermal compressibility ($\eta \kappa_T$),
    and the specific heat ($c_P$).
    Numerals beside each plot denote pressure values.
  }
  \label{fig:pcost}
\end{figure}

%% file: xptx.tex
\begin{figure}[t!]
  \includegraphics*[40mm,155mm][114mm,250mm]{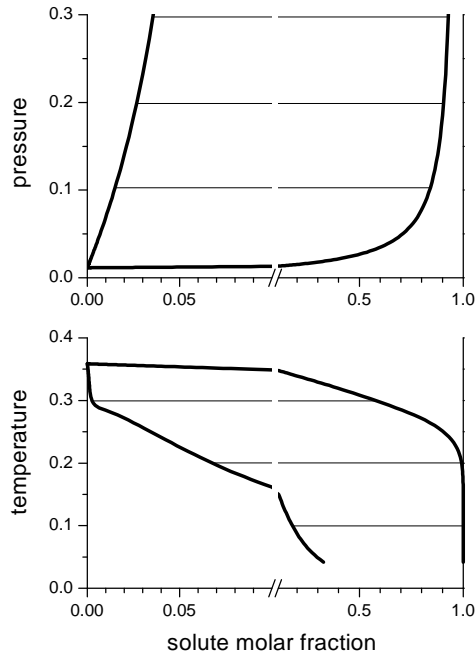}
  \caption
  {
    Constant temperature ($T/\eta=0.34$, top panel)
    and constant pressure ($P/\eta=0.015$, bottom panel) phase diagrams
    for a mixture of water ($\epsilon_\saf{ww}/\eta=0.3$, $c_\saf{w}=0.5$, $w=20$)
    with an ideal inert solute
    ($\epsilon_\saf{ws}=\epsilon_\saf{ss}=0$, $c_\saf{s}=0$).
    Thick lines denote pure phase boundaries;
    thin lines connect coexisting phases
    and denote phase-separated regions.
  }
  \label{fig:xptx}
\end{figure}

%% file: tdxs.tex
\begin{figure*}[t!]
  \includegraphics*[40mm,155mm][194mm,250mm]{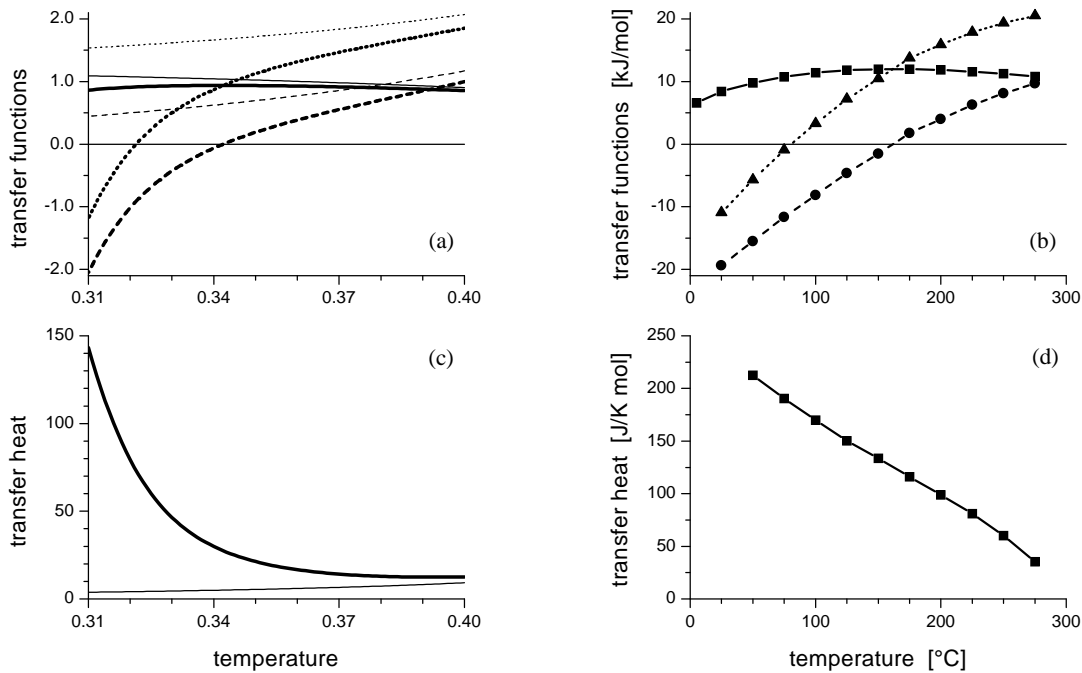}
  \caption
  {
    (a)~Transfer energy functions ($E/\eta$) vs temperature ($T/\eta$)
    for an ideal inert molecule in water at liquid-vapor coexistence:
    $E = \Delta g^*_\saf{s}$ (solid line),
    $E = T\Delta s^*_\saf{s}$ (dashed line),
    and $E = \Delta h^*_\saf{s}$ (dotted line).
    (b)~Corresponding experimental data
    for solvation of argon into water~\cite{BenNaim1987}.
    (c)~Transfer heat capacity ($\Delta {c_P}^*$) vs temperature
    ($T/\eta$) for an ideal inert molecule.
    (d)~Corresponding experimental data
    for argon~\cite{BenNaim1987}.
    Thin lines in (a) and (c) denote transfer quantities
    for nonbonding water.
  }
  \label{fig:tdxs}
\end{figure*}

%% file: tdxw.tex
\begin{figure*}[t!]
  \includegraphics*[40mm,195mm][194mm,250mm]{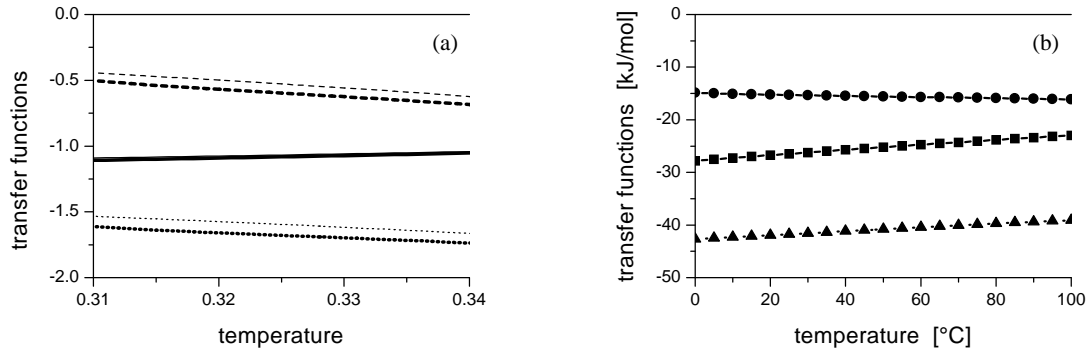}
  \caption
  {
    (a)~Transfer energy functions ($E/\eta$) vs temperature ($T/\eta$)
    for a water molecule into pure liquid water at liquid-vapor coexistence:
    $E = \Delta g^*_\saf{w}$ (solid line),
    $E = T\Delta s^*_\saf{w}$ (dashed line),
    and $E = \Delta h^*_\saf{w}$ (dotted line).
    (b)~Corresponding experimental data~\cite{BenNaim1987}.
    Thin lines in (a) denote transfer energies for nonbonding water.
  }
  \label{fig:tdxw}
\end{figure*}

%% file: entconv.tex
\begin{figure*}[t!]
  \includegraphics*[40mm,195mm][194mm,250mm]{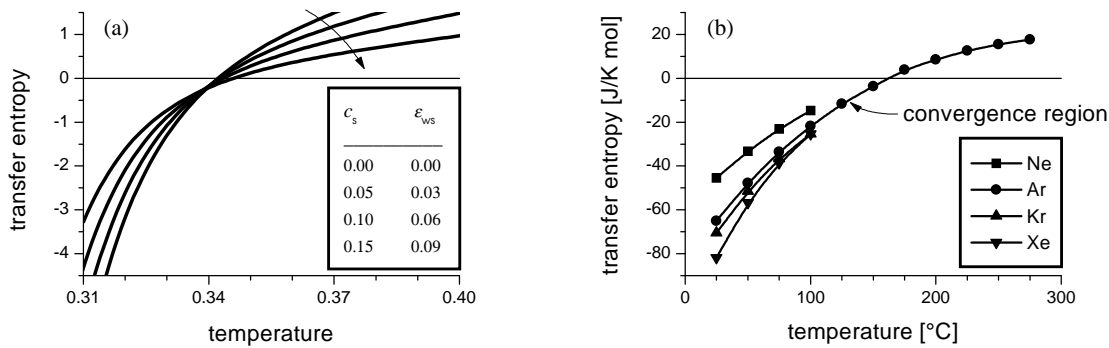}
  \caption
  {
    (a)~Transfer entropy ($\Delta s^*_\saf{s}$) vs temperature ($T/\eta$)
    for a solute molecule into pure liquid water at liquid-vapor
    coexistence, for different solute parameter values, with
    $\epsilon_\saf{ss} = 0$ and
    $\epsilon_\saf{ws}/\epsilon_\saf{ww} = c_\saf{s}/c_\saf{w}$.
    The arrow denotes increasing values of $c_\saf{s}$ and
    $\epsilon_\saf{ws}$.
    (b)~Corresponding experimental data for noble gases~\cite{BenNaim1987}.
  }
  \label{fig:entconv}
\end{figure*}